\documentclass[prl,aps,superscriptaddress,a4paper,twocolumn]{revtex4-1} 

\usepackage{graphicx}  
\usepackage{amsmath,amsfonts,amssymb,amsthm,amscd} 
\usepackage{hyperref}
\usepackage{physics}
\usepackage[mediumspace,mediumqspace,grey,squaren]{SIunits}

\hypersetup{
colorlinks=true,
linkcolor=blue,
linktoc=page,
citecolor=blue}

\begin{document}
\sloppy
\title{Coherent merging of counter-propagating exciton-polariton superfluids}

\author{T. Boulier}
\affiliation{Laboratoire Charles Fabry, Institut d'Optique Graduate School, CNRS, Universit\'{e} Paris-Saclay, 91127 Palaiseau cedex, France}

\author{S. Pigeon}
\affiliation{Laboratoire Kastler Brossel, Sorbonne Universit\'{e}, CNRS, ENS-PSL Research University, Coll\`ege de France, 4 place Jussieu Case 74, F-75005 Paris, France.}
\affiliation{Centre for Theoretical Atomic, Molecular and Optical Physics,
Queen’s University Belfast, Belfast BT7 1NN, United Kingdom.}

\author{E. Cancellieri}
\affiliation{Department of Physics and Astronomy, University of Sheffield, Hicks Building, Hounsfield Road, Sheffield, S3 7RH, England.}

\author{P. Robin}
\affiliation{Laboratoire Kastler Brossel, Sorbonne Universit\'{e}, CNRS, ENS-PSL Research University, Coll\`ege de France, 4 place Jussieu Case 74, F-75005 Paris, France.}

\author{E. Giacobino}
\affiliation{Laboratoire Kastler Brossel, Sorbonne Universit\'{e}, CNRS, ENS-PSL Research University, Coll\`ege de France, 4 place Jussieu Case 74, F-75005 Paris, France.}

\author{Q. Glorieux}
\affiliation{Laboratoire Kastler Brossel, Sorbonne Universit\'{e}, CNRS, ENS-PSL Research University, Coll\`ege de France, 4 place Jussieu Case 74, F-75005 Paris, France.}

\author{A. Bramati}
\email{bramati@lkb.umpc.fr}
\affiliation{Laboratoire Kastler Brossel, Sorbonne Universit\'{e}, CNRS, ENS-PSL Research University, Coll\`ege de France, 4 place Jussieu Case 74, F-75005 Paris, France.}

\date{\today}

\begin{abstract}
We report the formation of a macroscopic coherent state emerging from colliding polariton fluids. Four lasers with random relative phases, arranged in a square, pump resonantly a planar microcavity, creating four coherent polariton fluids propagating toward each other. When the density (interactions) increases, the four fluids synchronise and the topological excitations (vortex or soliton) disappear to form a single quantum superfluid. 
\end{abstract}

\maketitle

Can several quantum fluids of light interact so strongly as to merge and behave as one? Phase synchronization across coupled oscillators is a behavior observed in various systems, both classical~\cite{bennett2002huygens} and quantum~\cite{wiesenfeld1996synchronization,defreez1990spectral,kaka2006mutual}. In the latter case, a rich situation arises when several condensates, possessing independent initial phases, are coupled together. Their complete synchronization into a well-defined and continuous phase signifies their merging, an effect recently observed in ultracold atoms experiments~\cite{aidelsburger2017merging,scherer2007vortex}. However this remains an open question for strongly dissipative photonic systems, especially in the limiting case of two dimensions. The hallmark of this phenomenon is the vanishing of the phase singularities that are present for weak interactions. Here we report the observation of such a synchronization in a strongly dissipative light-based system: exciton-polaritons.

Exciton-polaritons, or simply polaritons, are quasiparticles born from the strong coupling between light (cavity photons) and matter (excitons) in a semiconductor microcavity~\cite{Weisbuch92}. Polaritons are characterized by a low effective mass, inherited from their photonic component, and strong nonlinear interactions due to their excitonic part. They offer a great opportunity to revisit in solid-state materials fundamental concepts first explored in the context of atomic physics. Moreover, polaritonic systems are easily controllable by optical techniques and, due to their finite lifetimes, are ideal systems for studying out-of-equilibrium phenomena~\cite{RevModPhys.82.1489,Carusotto13}. In analogy with the atomic case~\cite{RevModPhys.71.463,PhysRevLett.86.4447}, condensation~\cite{Kasprzak06} and the superfluid behavior of polaritonic quantum fluids~\cite{Amo11} have been of great theoretical interest~\cite{PhysRevLett.93.166401,PhysRevLett.105.020602,PhysRevB.82.224512} and have been experimentally confirmed~\cite{Amo09b,Amo09a,sanvitto10}.
  
Recently, multi-pumps settings were explored to form polariton condensates by off-resonant excitation of spatially distinct areas~\cite{cristofolini13,tosi12}, and to study the collision of strongly interacting fluids in a resonant excitation regime~\cite{hivet2014interaction,Boulier15,goblot2016phase}. This allows for example the condensation of polaritons in a pump-free zone surrounded by the excitation spots~\cite{cristofolini13} or the formation of nonlinear collective excitations (dark solitons and vortices) with resonant driving~\cite{hivet2014interaction,Boulier15,goblot2016phase}. Interestingly, it is found that the density of such excitations diminishes as the interactions increase~\cite{goblot2016phase} and that the strong interactions regime can lead to the merging of vortex-antivortex (V-AV) pairs~\cite{cancellieri14}, paving the way to a perfect merging of distinct Bose gases, as reported here. 


\begin{figure*}
\begin{center}
\includegraphics[width=0.9\textwidth]{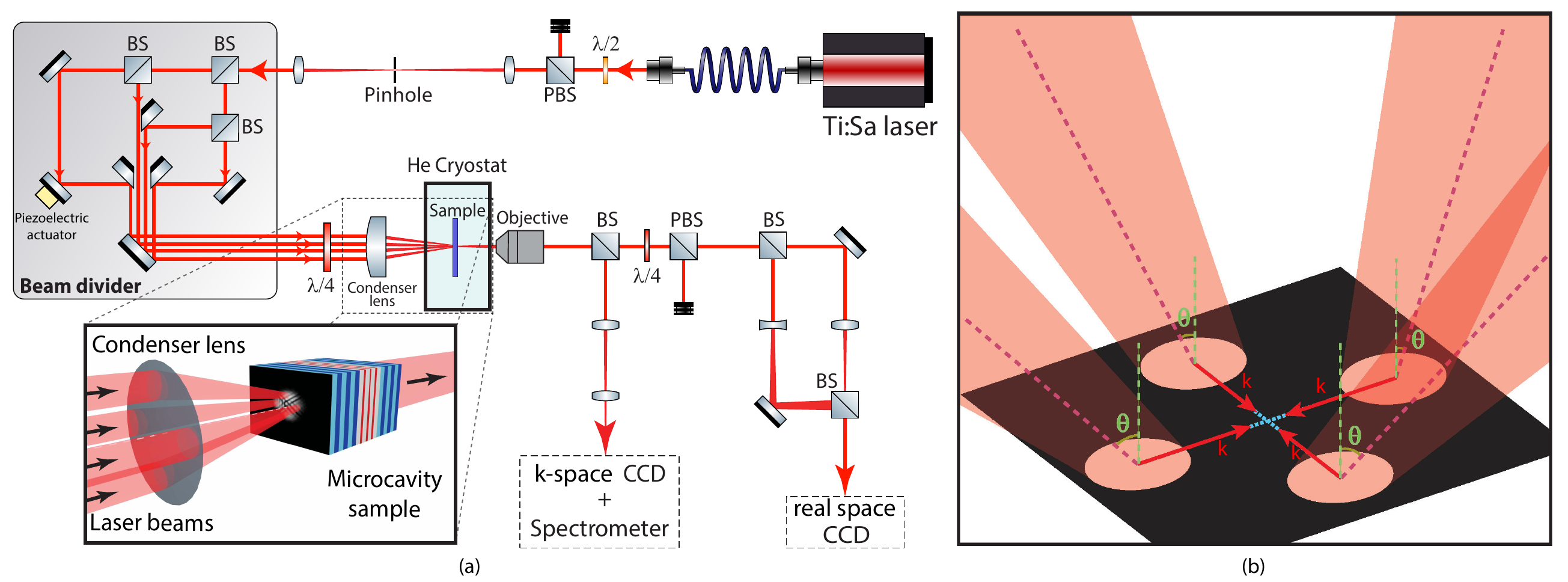}
\caption{\textbf{Pumping configuration.} Four coherent beams from the same laser are focalized on the sample to form a square geometry. The relative phases of the pumps are fixed to random values. One pump is reflected on a piezoelectric-actuated mirror, which allows to vary its phase. The incident angle~$\theta$ controls the fluid planar momentum $\hbar k$. Here the four fluids propagate towards the center of the square geometry with the same momentum. The pumps are separated by~$R=\unit{20}{\micro\metre}$.}
\label{Setup4pumps}
\end{center}
\end{figure*}

In the present work, we focus on achieving the complete vanishing of all interference and the annihilation of all phase singularities inside a two-dimensional, square geometry four-pumps system. We observe the merging of four polaritons populations, each initially defined by its distinct group velocity and phase, into a single coherent population possessing a smoothly varying velocity field. We study the conjugate momentum space through far-field imaging, which shows an unexpected pattern in the superfluid regime. Strikingly, the phase of each fluid, initially imposed by the driving lasers, is modified as a consequence of the interactions. This is opposite to the resonant pumping paradigm used in experiments such as Ref.~\cite{goblot2016phase}, by which the pump fixes the boundary condition for the wavefunction. To model our results, we develop two distinct theoretical approaches, one numerical and the other analytical. Both are in good agreement with the observations and show a genuine out-of-equilibrium phase synchronization between four colliding Bose gases. 

\emph{Experimental setup --}
The sample is a planar microcavity formed by three GaAs/InGaAs quantum wells sandwiched between two Bragg mirrors. A more detailed description of the sample and the experimental setup are available in Ref.~\cite{BoulierTilt}. Polaritons are resonantly pumped with a continuous wave laser, frequency-locked to an optical cavity at~$\unit{837}{\nano\meter}$. We can choose the photon-exciton energy detuning at normal incidence~$\delta=E_\text{ph}\left(\mathbf{0}\right)-E_\text{exc}\left(\mathbf{0}\right)$ (where~$\mathbf{k}$ is the in-plane polariton wavevector, $E_\text{ph}(\mathbf{k})$ and $E_\text{exc}(\mathbf{k})$ are respectively the photon and exciton energies) thanks to a wedge in the optical cavity. All the measurements presented here were taken at~$\delta=\unit{+0.5}{\milli\electronvolt}$. 

The laser beam is separated into four arms of equal intensity, focalized on the sample with a condenser lens to form a square geometry (see~Fig.~\ref{Setup4pumps}). Their position in k-space is chosen so that polaritons propagate precisely towards the square center, such that they have exact opposite direction (unlike in~\cite{Boulier15}). To avoid overlaps between different pumps~\cite{vladimirova10}, the laser beam is spatially filtered with a pinhole prior to division into the pump beams, such that the Gaussian tail is blocked. We use circularly polarized light to pump a single pseudo-spin state in all fluids. Here the resonant pumping configuration allows the fine-tuning of the polariton density, and therefore interactions, but does not generate an excitonic reservoir. This is in contrast with the observations performed with an out-of-resonance setup~\cite{tosi12,cristofolini13}. The four in-plane wavevectors are chosen with the same norm~$k_p=\frac{2\pi}{\lambda} \sin \theta$, by choosing the same angle of incidence~$\theta$ for all pumps.

The cavity photoluminescence is collected in transmission geometry with a microscope objective to be analyzed both in real and momentum space. It is then imaged onto a CCD camera to record the intensity distribution. The polariton phase is measured with a modified Mach--Zehnder interferometric setup: a beam splitter divides the real space image into two parts, one of which is widely expanded to generate a flat phase reference beam, which is used to make an off-axis interference pattern. The actual phase map is then numerically retrieved by Fourier analysis. All experiments are performed at a temperature of $\unit{10}{\kelvin}$, with liquid helium cryogenics.

\emph{Model --}
To better understand this experiment, we use a mean-field model of the polariton field. Only considering the lower polariton branch, we numerically solve the 2D driven-dissipative Gross--Pitaevskii equation to access the time evolution of the system. Taking $\psi \equiv \psi(\mathbf{r}) = \langle\hat{\psi}(\mathbf{r})\rangle$, the complex mean value of the polariton field operator $\hat{\psi}(\mathbf{r})$, we solve
\begin{equation}
i\hbar \frac{\partial \psi}{\partial t}=\left(-\frac{\hbar^2\nabla^2}{2m^*}-\frac{i\hbar \gamma}{2}+ g \vert \psi\vert^2 \right)\psi+F(\mathbf{r})e^{i \frac{\Delta t}{\hbar}},
\label{GP}
\end{equation}
where $m^*$ is the polariton effective mass (typically~$10^{-4}$ electron mass), $\gamma$ is the decay rate deduced from the polariton lifetime ($\unit{12}{\pico\second}$ for our sample), and $g=\unit{5}{\micro\electronvolt \micro\meter^2}$ is the polariton-polariton interaction strength. $\Delta$ is the energy detuning between the pump laser and the polariton resonance. In this experiment, $\Delta$ is fixed to $\unit{0.3}{\milli\electronvolt}$.
Here, $F(\mathbf{r})=\sum_{i=1}^4\sqrt{I(\mathbf{r}-\mathbf{r}_i)}e^{i\mathbf{r.k}_i+\phi_i}$, where $I(\mathbf{r})$ is a pump intensity profile after the pinhole, $\phi_i$ and $\mathbf{r}_i$ are respectively a random phase and the position on the sample associated to the pumping spot $i$ and $\mathbf{k}_i \in \{ \binom{k_p}{0},\binom{-k_p}{0},\binom{0}{k_p},\binom{0}{-k_p}  \}$ corresponding in-plane wavevector. Each pumping spot has a diameter of $\unit{12}{\micro\meter}$. Direct comparison with the experiment is then performed by extracting the density $n=\abs{\psi}^2$ and phase $\arg(\psi)$ after steady-state is numerically reached. 

Here we present our results in two extreme regimes, at low density (negligible interactions, $gn \ll \hbar\gamma$) and at high density (strong repulsive interactions, $gn \gg \hbar\gamma$):

\begin{figure}[t]
	\includegraphics[width=0.45\textwidth]{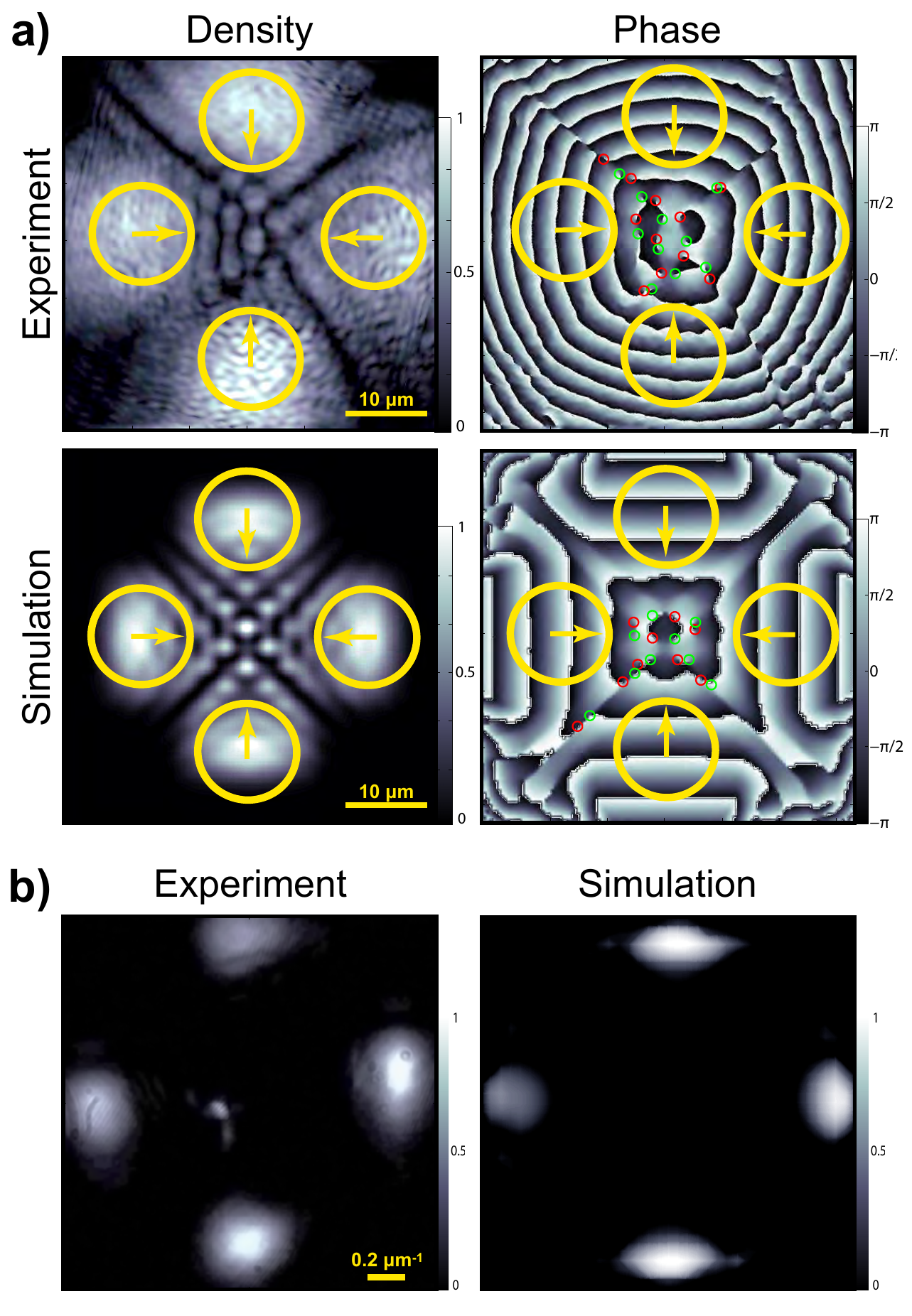}
\caption{\textbf{Linear regime at low polariton density} - a) Experimental (up) and numerical (down) results of real space density and phase distribution at low excitation power ($\unit{10}{\milli\watt}$ for each pump). An interference pattern visible between the contra-propagating pumps contains vortices (green and red circles depending on their circulations). The yellow circles delimit the pumps while the yellow arrows show the polariton flow. - b) k-space emission in the low interactions regime. Only the four pumped momenta are visible (the spot near the center of the experimental maps is a parasite reflection from the experimental setup). The intensity scale is linear and normalized to the peak intensity.}
	\label{4PumpLowDRspace}
\end{figure}

\emph{Low density regime --}
When the density is low enough, the interaction energy is negligible. The polaritons coherence, inherited from the pumps, simply leads to interference between the four non-interacting fluids (see~Fig.~\ref{4PumpLowDRspace}). This regime is analogous to the purely photonic, linear optics case. The square cells of this interference pattern is typical of a four-beam interference pattern~\cite{hivet2014interaction}. Note that the experiment pattern in Fig.~\ref{4PumpLowDRspace} is not a perfect square pattern, due to the sample inhomogeneity and the pumps imperfect geometry. We also see that the interference nodes often contain phase singularities. The pumps relative phases $\phi_i$ are set randomly, which gives rise to phase dislocations between pumps visible in the interferogram of Fig.~\ref{4PumpLowDRspace}a.

Fig.~\ref{4PumpLowDRspace}b shows the momentum space, i.e. the distribution of polaritons wavevectors, in the low density regime. Only the pumps wavevectors are present in this non-interacting regime. This is expected as each population propagates without significant scattering event, except on the cavity natural disorder~\cite{hivet13, Leyder07}. The clear separation of each wavevector population justifies considering the system as the sum of four distinct polariton fluids.

\begin{figure}[t]
\includegraphics[width=0.45\textwidth]{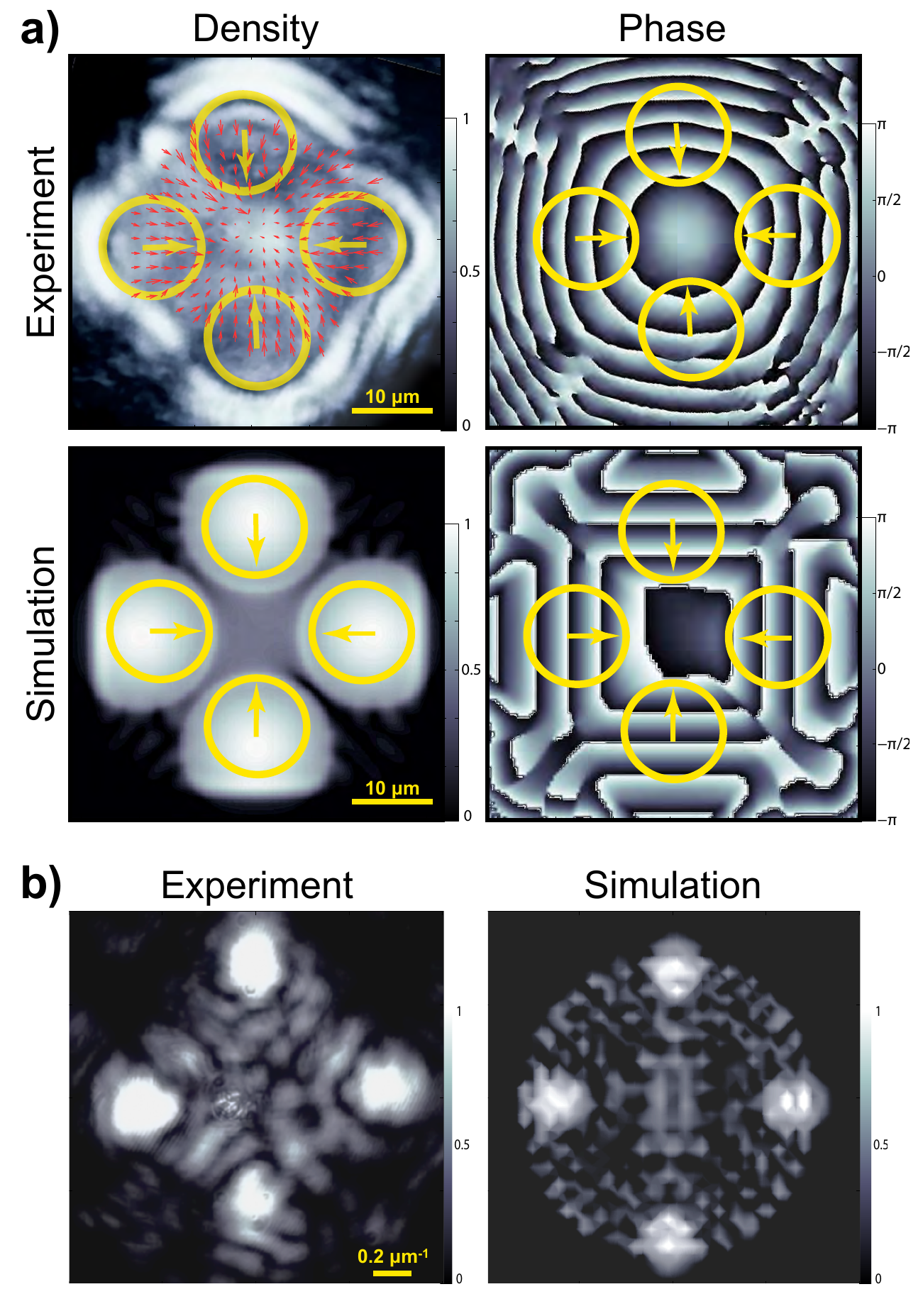}
\caption{\textbf{Nonlinear regime at high polariton density} - a) Experimental (up) and numerical (down) real space density (left) and phase (right) in the high density regime. Here no interference pattern is visible as the four superfluids merge coherently into one. The red arrows show the velocity field. - b) k-space emission in the high density regime. The intensity scale is linear and normalized to the peak intensity. Here the interactions trigger the emergence of a population for low wavevectors.}
\label{4PumpHigDRspace}
\end{figure}

\emph{High density regime --}
Previous works showed the transition from a linear interference pattern to hydrodynamic soliton physics~\cite{goblot2016phase,hivet2014interaction}. Here we focus on high interactions (a pump power of about $\unit{100}{\milli\watt}$ for each pump). We consistently observe that the phase is modified so as to form a single coherent superfluid, with no phase jump nor singularity, as visible on Fig.~\ref{4PumpHigDRspace}a. The interaction energy is always locally much larger than the kinetic energy in this regime, and is responsible for this macroscopic phase synchronization. The numerical simulations, presented next to the experimental data in Fig. \ref{4PumpHigDRspace}, confirm this smooth, coherent phase of the resulting fluid.

\begin{figure}[t]
\includegraphics[width=0.45\textwidth]{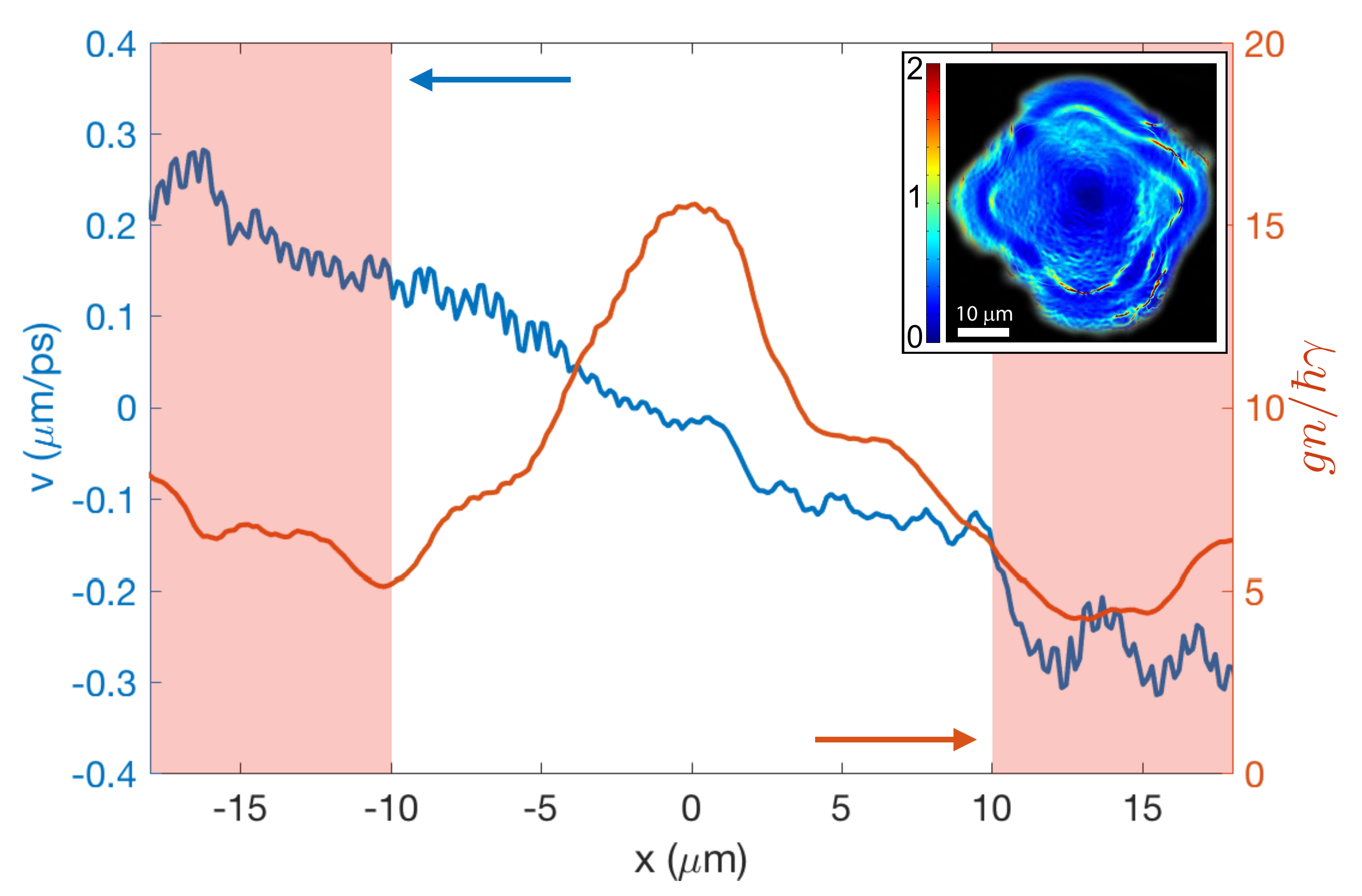}
\caption{\textbf{Spatial cut of polariton fluid properties} Velocity (in units of $c_s$, blue line) derived from the phase gradient along the x-axis, and interaction energy shift (orange line) derived from the density. Inset: full 2D map of the Mach number $M=\abs{v/c_s}$.
Within both pump areas, indicated by the light red regions, no significant energy shift is observed whereas in-between them the velocity varies linearly. 
}
\label{4PumpHigDVelo}
\end{figure}

Note that since the resonant pumping does not populate exciton reservoir unlike non-resonant driving, no confining potential is present. This can be seen in Fig.~\ref{4PumpHigDVelo} (orange curve), where the local energy shift due to polariton interaction present a local maximum in the central region. Therefore one can rule out confinement effects to explain the merging of the four superfluids, as opposed to recent experiments with polariton condensates~\cite{cristofolini13}. 

This is all the more striking as it is independent from the pumps relative phases. A mirror used to reflect one of the pumps was mounted on a piezoelectric transducer, so that one of the pump can have its phase varied between $0$ and $2\pi$ relative to the others. The resulting superfluid remains unchanged when the relative phase of a single pump spot is modified. Likewise, the emission systematically showed a smooth phase across the pumps for any relative pump phase. Therefore no phase domain nor density defect can be detected: the system remains as a single superfluid, unlike in the linear case. Through high interactions, polaritons gain a well-defined collective phase similarly to results obtained with cold atom quantum fluids \cite{aidelsburger2017merging,scherer2007vortex}.

In k-space, the signal shows a more complex pattern than in the linear regime (see Fig.~\ref{4PumpHigDRspace}b). In accordance with the observed velocity field, a signal is now present for wavevectors lower than that of the pumps. This situation is unlike the weakly interacting regime: the presence of a wavevectors continuum does not permits the clear-cut identification of four polariton populations, which confirms the need to consider the system as single fluid.

As an additional check, the real-space position of a specific wavevector was studied using a mobile pinhole in the detection system. A single wavevector component is selected and the corresponding signal is observed on the density picture. We found that the $k=\unit{0}{\micro\meter^{-1}}$ emission at the center of Fig.~\ref{4PumpHigDRspace}b corresponds to the center of the system. This is in good agreement with the velocity extracted from the phase profile and shown as a red vector field in Fig.~\ref{4PumpHigDRspace}a: Polaritons from each pump lose velocity as they approach the center. This behavior, only visible at high densities, is consistent with the merging of the four fluids. For two contra-propagating fluids a smooth phase implies the continuous transition from $-k_p$ to $k_p$ (since the direction changes), which in turns implies a zero-velocity population at the center, as is also visible in the velocity cut in Fig.~\ref{4PumpHigDVelo} (blue curves).

We also experimentally checked the total energy (kinetic plus interaction) of the superfluid remains constant during propagation. The loss of kinetic energy near the center is compensated by a larger interaction energy due to the higher density. This observation is not trivial for a highly dissipative quantum fluid. 

\emph{Conclusion}
 We designed a scheme that allows the study of two-dimensional collisions of several polariton Bose gases, in the regime of high interactions. We demonstrate the formation of macroscopic, coherent, non-homogeneous superfluid flow of polaritons, free from topological excitations. While in the linear regime interferences appear and phase singularities are clearly visible, the vanishing of all singularities happens at sufficiently high polariton density, as expected from previous studies~\cite{hivet2014interaction, cancellieri14,goblot2016phase}. This enables the disappearance of all defects (phase and density) in real-space, and the the appearance of a quasi-continuum of low wavevector states in the reciprocal space.\\* 
This result shows the complexity of quantum fluids of light, and offers a new insight on the collective behavior of interacting polariton superfluids. Our results are expected to stimulate more studies of multi-pump systems, where exotic dispersion curves are predicted. It also opens the door to new, complex multi-pump experiments with which one could precisely manipulate the superfluid shape and supercurrents. Such a scheme could, for example, allow the study of fundamental problems such as the Kibble-Zurek mechanism, in a similar fashion to current cold atoms BEC experiments~\cite{aidelsburger2017merging}.

\begin{acknowledgments}
We acknowledge the financial support of the ANR projects Quandyde, QFL, and C-FLigHT. T.B. acknowledges the support of the European Marie Sk\l{}odowska-Curie Actions (H2020-MSCA-IF-2015 Grant 701034).
\end{acknowledgments}



\end{document}